\documentclass[twocolumn,floatfix,superscriptaddress,a4paper,
               showpacs,showkeys,nofootinbib,reprint,prl]{revtex4}
\usepackage{epsfig}
\usepackage{latexsym}
\usepackage{xspace}
\usepackage{hyperref}
\usepackage[latin2]{inputenc}
\usepackage{indentfirst}
\usepackage{enumerate}
\usepackage{color}

\usepackage{amsmath}
\usepackage{amssymb}
\usepackage[english]{babel}
\usepackage{url}
\newcommand{\eq}[1]{\begin{align} #1 \end{align}}

\begin{document}

\title{Surprisingly large uncertainties in 
temperature extraction from
thermal
fits to hadron yield data at LHC}
\author{Volodymyr Vovchenko}
\affiliation{
Frankfurt Institute for Advanced Studies, Goethe Universit\"at Frankfurt,
D-60438 Frankfurt am Main, Germany}
\affiliation{
Taras Shevchenko National University of Kiev, 03022 Kiev, Ukraine}
\affiliation{
Institut f\"ur Theoretische Physik,
Goethe Universit\"at Frankfurt, D-60438 Frankfurt am Main, Germany}
\author{Horst Stoecker}
\affiliation{
Frankfurt Institute for Advanced Studies, Goethe Universit\"at Frankfurt,
D-60438 Frankfurt am Main, Germany}
\affiliation{
Institut f\"ur Theoretische Physik,
Goethe Universit\"at Frankfurt, D-60438 Frankfurt am Main, Germany}
\affiliation{
GSI Helmholtzzentrum f\"ur Schwerionenforschung GmbH, D-64291 Darmstadt, Germany}

\date{\today}

\begin{abstract}
The conventional hadron-resonance gas (HRG) model with the Particle Data Group (PDG) hadron input, full chemical equilibrium, and the 
hadron type 
dependent eigenvolume interactions is employed to fit the hadron mid-rapidity yield data of ALICE Collaboration for the most central Pb+Pb collisions.
For the case of point-like hadrons the well-known fit result $T = 154 \pm 2$~MeV is reproduced. 
However, the situation changes if hadrons have different eigenvolumes.
In the case when all mesons are point-like while all baryons have an effective hard-core radius of 0.3 fm
the $\chi^2$ temperature dependence of the $\chi^2$ has a broad minimum in the temperature range of $155-210$ MeV, with fit quality comparable to the $T \sim 155$ MeV minimum in the point-particle case.
Very similar result is obtained when only baryon-baryon eigenvolume interactions are considered, with eigenvolume parameter taken from previous fit to ground state of nuclear matter. 
Finally, when we apply the eigenvolume corrections with mass-proportional eigenvolume $v_i \sim m_i$, fixed to particular proton hard-core radius $r_p$, we observe a second minimum in the temperature dependence of the $\chi^2$, located at the significantly higher temperatures.
For instance, at $r_p = 0.5$~fm the fit quality is better than in the point-particle HRG case in a very wide temperature range of $170-320$ MeV, 
which gives an uncertainty in the temperature determination from the fit to the data of 150~MeV.
These results show that thermal fits to the heavy-ion hadron yield data are
very sensitive to
the modeling of the short-range repulsion eigenvolume between hadrons, and that 
chemical freeze-out temperature 
can be extracted
from the LHC hadron yield data
only with sizable uncertainty.
\end{abstract}



\maketitle

\section{Introduction}
Thermodynamic models are valuable tools in the modern-day high energy physics, and have long been employed to estimate the temperatures reached in the relativistic heavy-ion collisions~\cite{SOG1981, Molitoris1985,Hahn1987}.
The hadron-resonance gas model and its modifications have been rather successfully used
to extract the chemical freeze-out properties of matter created in heavy-ion collisions,
by fitting the rich data on mean hadron multiplicities in various experiments,
ranging from the low energies at Bevalac and SchwerIonen-Synchrotron (SIS) to the highest energy of
the Large Hadron Collider
(LHC) at CERN~\cite{CleymansSatz,CleymansRedlich1998,CleymansRedlich1999,
Becattini2001,BraunMunzinger2001,Becattini2004,ABS2006,ABS2009}.
The HRG model is commonly used to describe the low-temperature part of QCD but
it has also been suggested that the HRG contributes appreciably to the total QCD pressure also at higher temperatures~\cite{Biro}.
In a realistic HRG model one has to take into account the attractive and
repulsive interactions between hadrons.
It has been argued~\cite{DMB}, that the
inclusion into the model of all known resonances as free non-interacting (point-like) particles allows to
effectively model the attraction between hadrons.
This formulation, a multi-component point-particle gas of all known hadrons and resonances, is presently the most commonly used one in the thermal model analysis. At the highest collision energies available at the LHC the asymmetry in production of particles and anti-particles becomes vanishingly small, which implies that the (baryo)chemical potential $\mu_B$ is close to zero, and the hadron yield ratios are determined by a single parameter, the chemical freeze-out temperature $T$.
\footnote{We do not consider here possible under- or over-saturation of the light and/or strange quarks which would improve the data description but would also introduce additional parameters~(see, e.g., Ref.~\cite{LetRaf,MHBQ}).}

In order to take
the short-range repulsive interactions
into account, the eigenvolume (EV) correction of the van der Waals type was included into hadronic equation of state first in Refs.~\cite{HagedornEV1,Gorenstein1981,Kapusta1983}. The most commonly used EV model was formulated in Ref.~\cite{Rischke1991}, and later extended to the multi-component case in Ref.~\cite{Yen1997}.
The eigenvolume HRG model has been successfully compared with the lattice QCD data at lower temperatures~\cite{EV-latt-1,EV-latt-2}, and a combined HRG + perturbative QCD fit to the lattice data was shown to prefer hadrons with finite eigenvolume over point-like hadrons~\cite{EV-latt-3}. 

In the multi-component EV model~\cite{Yen1997} one has to solve the transcendental equation for the pressure, which reads as
\eq{\label{eq:Pex}
P(T, \mu) = \sum_{i} \, P^{\rm id}_i (T, \mu_i^*),
}
where the sum goes over all hadrons included in the set, $P^{\rm id}_i (T, \mu_i^*)$ is the pressure of the ideal (point-like) Fermi or Bose gas at the corresponding temperature and chemical potential, and
$\mu_i^* = \mu_i - v_i \, P(T, \mu)$. The $v_i$ is the eigenvolume parameter\footnote{
Hadron eigenvolume parameter $v_i$ and effective its hard-core radius $r_i$ are related to each other as $v_i = 4 \cdot 4 \pi r_i^3/3$. 
This relation
can be rigorously obtained for
the low-density gas of hard spheres (see e.g., Ref.~\cite{LL}).
We note that the true value of $r_i$ found in interaction potential may be different due to quantum-mechanical corrections~\cite{KGSG,Typel:2016srf}.
}
of the hadron species $i$
, and the number density of these species can be calculated as
\eq{\label{eq:nex}
n_i(T, \mu) = \frac{n_i^{\rm id} (T, \mu_i^*)}{1 + \sum_j v_j n_j^{\rm id} (T, \mu_j^*)}.
}
In our analysis we include the established strange and non-strange hadrons listed in the Particle Data Tables~\cite{pdg}, along with their decay branching ratios. This includes mesons up to $f_2(2340)$ and (anti)baryons up to $N(2600)$.
The finite width of the resonances is taken into account in the usual way, by adding the additional integration over their Breit-Wigner shapes in the point-particle gas expressions.
Additionally, we also consider the case when the finite width of the resonances is neglected.
The feed-down from decays of the unstable resonances to the total hadron yields is included in the standard way.

The thermal fits within the chemical equilibrium HRG model formulation performed within the different codes have consistently yielded the chemical freeze-out temperature of $T \simeq 155$~MeV~\cite{SHARE,SABR2014,Floris,Becattini2014} for the most central Pb+Pb collisions at $\sqrt{s_{\rm NN}} = 2.76$~TeV. 
These analyses, however, either neglect the short-range repulsive interactions between the hadrons, or use the same eigenvolumes for all hadrons.
The eigenvolume corrections can significantly reduce the densities~\cite{Begun2013,mf-2014}, and, thus, increase the total system volume at the freeze-out compared to a point-particle gas at the same temperature and chemical potential. This correction, however, has a very small effect on the hadron yield ratios, and, thus, on the extracted freeze-out temperature, in the cases when the eigenvolumes of different hadrons are (nearly) identical. On the other hand, if one considers hadrons with the different hard-core radii, then the 
hadron yield ratios change,
and the extracted
temperature (and the chemical potential) may become different
compared to the point-particle case~\cite{Gorenstein1999}.

\section{Scenarios for eigenvolume interactions}

The values of hadron eigenvolumes are presently not well constrained and are the source of a systematic uncertainty in the HRG model.
In order to study the sensitivity of the thermal fits
to that uncertainty 
we employ 
three different parametrizations for the hadron EV interactions.

\begin{enumerate}
\item In the first case we assume that all mesons are point-like, i.e. $v_M = 0$,
and that all baryons have a fixed finite EV $v_B > 0$.
Note that in this case mesons and baryons still ``see'' each other: the point-like mesons cannot penetrate into the finite-sized baryons.
For the effective hard-core radius of baryon we use the value $r_B = 0.3$~fm. 
This choice is motivated by a successful comparison of this model with the lattice QCD data for the pressure reported in Ref.~\cite{EV-latt-1}: 
the lattice data was described up to at least $T=190$~MeV. 
Our calculations suggest that lattice pressure is described fairly well even up to 250~MeV within this model.

\item
In the second case we employ
a bag-model inspired parametrization with the hadron eigenvolume proportional to its mass through a bag-like constant, i.e. 
\eq{\label{eq:BagEV}
v_i = m_i / \varepsilon_0.
}
Such eigenvolume parametrization had been obtained for the heavy Hagedorn resonances, and was used to describe their thermodynamics~\cite{HagedornEV1,Kapusta1983} as well as their
effect on the particle yield ratios~\cite{NoronhaHostler2009}.
In the absence of the detailed knowledge regarding the different hadron-hadron interactions,
we adopt here this parametrization for all hadrons.
Note that the HRG with eigenvolume corrections given by \eqref{eq:BagEV} have recently been used to model the hadronic part of the crossover QCD equation of state,
which compares favorably to the lattice data~\cite{EV-latt-3}.
It was mentioned in the Ref.~\cite{Becattini2006} that this kind of parametrization leads to the increase of the freeze-out temperature, but that it does not necessarily entail an improvement of the fit quality. 
We shall study this question here for the LHC energies. 
We also note that the eigenvolume for the resonances with the finite width is assumed to be constant for each resonance, and is determined by its pole mass.

\item In the third case we include only baryon-baryon and antibaryon-antibaryon EV interactions and neglect the EV interactions for baryon-antibaryon, meson-baryon, and meson-meson pairs. In such case the system consists of three independent
sub-systems: non-interacting mesons, EV baryons, and EV antibaryons. In this case meson densities are given by the ideal gas relations.
In order to calculate the densities of (anti)baryons Eqs.~\eqref{eq:Pex} and \eqref{eq:nex} are solved, separately for baryons and antibaryons. The eigenvolume parameter value $v_B$ of the baryon-baryon interaction is fixed from the fit to the ground state of nuclear matter within the van der Waals equation performed in Ref.~\cite{Vovchenko:2015vxa}. The resulting value is $v_B = 3.42$~fm$^3$, which corresponds to the effective hard-core radius of $r_B \simeq 0.59$~fm. Our consideration of this scenario is motivated by the successful recent comparison of such model with lattice QCD, which was performed for many observables~\cite{Vovchenko:2016rkn}.\footnote{This model is denoted as EV-HRG in Ref.~\cite{Vovchenko:2016rkn}. In the present work we neglect the van der Waals attraction between baryons.}
Many improvements in the description of lattice data over the point-particle HRG were reported.
\end{enumerate}

\section{Results of the thermal fits}

We perform the thermal fit to the midrapidity yields of the charged pions, charged kaons, (anti)protons, $\Xi^-$, $\Xi^+$, $\Omega$, $\bar{\Omega}$, $\Lambda$, $K^0_S$, and $\phi$, measured by the ALICE collaboration in the 0-5\% most central Pb+Pb collisions at $\sqrt{s_{\rm NN}} = 2.76$~TeV~\cite{ALICEpiKp,ALICELamb,ALICEhyper,ALICEphi}. Note that the centrality binning for $\Xi$ and $\Omega$ hyperons is different from the other hadrons. Thus, we take the midrapidity yields of $\Xi$ and $\Omega$ in the $0-5$\% centrality class from Ref.~\cite{Becattini2014}, where they were obtained using the interpolation procedure.
The data used in thermal fits is listed in Table~\ref{tab:data}.
Our own implementation of the HRG model is used in the analysis. Additionally, we reproduce the obtained results within the publicly-available THERMUS-2.3 package~\cite{THERMUS}, which includes the implementation of the eigenvolume model given by the Eqs.~\eqref{eq:Pex} and \eqref{eq:nex}.

\begin{table}
 \caption{The hadron midrapidity yields for $0-5$\% most central Pb+Pb collisions measured by the ALICE collaboration and used in the thermal fits throughout this work.} 
 \centering                                                 
 \begin{tabular}{c|c|c}                                   
 \hline
 \hline
 Particle & Measurement ($dN/dy$) & Reference \\
 \hline
 $\pi^+$       & $733 \pm 54$ & \cite{ALICEpiKp} \\
 $\pi^-$       & $732 \pm 52$ & \cite{ALICEpiKp} \\
 $K^+$         & $109 \pm 9$  & \cite{ALICEpiKp} \\
 $K^-$         & $109 \pm 9$  & \cite{ALICEpiKp} \\
 $p$           & $34  \pm 3$  & \cite{ALICEpiKp} \\
 $\bar{p}$     & $33  \pm 3$  & \cite{ALICEpiKp} \\
 $\Lambda$     & $26  \pm 3$  & \cite{ALICELamb} \\
 $\Xi^-$       & $3.57  \pm 0.27$  & \cite{ALICEhyper}, \cite{Becattini2014} \\
 $\Xi^+$       & $3.47  \pm 0.26$  & \cite{ALICEhyper}, \cite{Becattini2014} \\
 $\Omega + \bar{\Omega}$       & $1.26  \pm 0.22$  & \cite{ALICEhyper}, \cite{Becattini2014} \\
 $K^0_S$       & $110  \pm 10$  & \cite{ALICELamb} \\
 $\phi$        & $13.8  \pm 0.5 \pm 1.7$  & \cite{ALICEphi} \\
\hline
\hline
 \end{tabular}
\label{tab:data}
\end{table}

\begin{figure}[!t]
\centering
\includegraphics[width=0.49\textwidth]{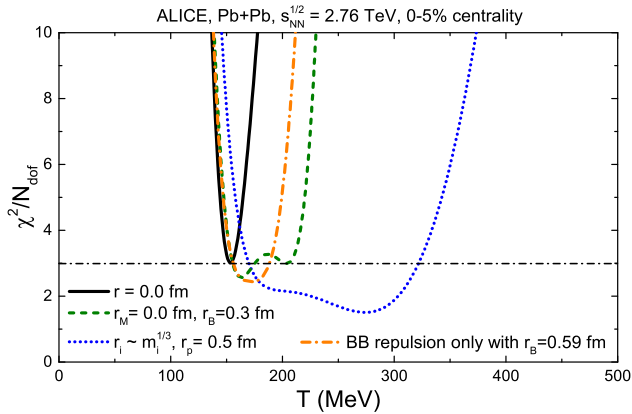}
\caption[]{(Color online)
The temperature dependence of $\chi^2 / N_{\rm dof}$ of fit to ALICE data on hadron yields in 0-5\% most central Pb+Pb collisions at 2.76~TeV within the point-particle HRG model (solid black curve),
the two-component eigenvolume HRG model with point-like mesons and baryons of fixed size (dashed green line),
the bag-like eigenvolume HRG model (dotted blue line),
and the model with only baryon-baryon and antibaryon-antibaryon eigenvolume interactions (orange dash-dotted line).
In the first case the effective hard-core radius of baryons is fixed at $r_B = 0.3$~fm,
in the
second case the bag-like constant in Eq.~\eqref{eq:BagEV} is fixed in order to reproduce the effective hard-core proton radius of 0.5 fm,
and in the third case effective hard-core radius of the baryon-baryon EV interaction is fixed to 0.59~fm, as explained in the text.
}\label{fig:chi2-vs-T}
\end{figure}

Figure~\ref{fig:chi2-vs-T} shows the dependence of the $\chi^2/N_{\rm dof}$ of the fit on the temperature for four cases: 
the gas of point-particle hadrons, i.e. for $v_i=0$, 
and for the three described scenarios with eigenvolume interactions between hadrons.
At each temperature the only remaining free parameter, namely the system volume (radius) 
per unit slice of rapidity,
is fixed to minimize the $\chi^2$ at this temperature.

All considered cases give drastically different pictures. 
For the point-particle HRG we obtain a narrow minimum around $T \simeq 154$~MeV with $\chi^2/N_{\rm dof} \simeq 30.1/10$.
If the finite width of resonances is neglected one gets a smaller $\chi^2/N_{\rm dof} \simeq 20.7/10$ with essentially the same value of temperature. The extracted temperature is consistent with the previous studies dealing with fits to the ALICE hadron yields~\cite{SABR2014,Floris}.
The presence of the finite eigenvolume generally increases the freeze-out temperature and improves the fit quality. 
The eigenvolume first case, the one with point-like mesons ($r_M = 0$~fm) and baryons of finite size ($r_B = 0.3$~fm) yields a broad (double) minimum at $155-210$~MeV in the temperature dependence of the $\chi^2$ (see dashed green line in Fig.~\ref{fig:chi2-vs-T}). The global minimum is located at $T \simeq 163$~MeV with $\chi^2/N_{\rm dof} \simeq 25.6/10$. In general, the fit quality stays comparable to the point-particle case in the whole $155-210$~MeV temperature range.
We note again that the thermodynamic functions of the eigenvolume HRG given by the Eqs.~\eqref{eq:Pex} and \eqref{eq:nex} with
$r_M = 0$~fm and $r_B = 0.3$~fm were compared in Ref.~\cite{EV-latt-1} to the lattice QCD data
at temperatures up to $T=190$~MeV, and a fairly good agreement was obtained. 
While we do not necessarily require that the lattice gauge theory can be directly projected on the heavy ion data we note that our $T=155-210$~MeV ALICE fit with $r_M = 0$~fm and $r_B = 0.3$~fm is consistent with the lattice data.
The fit result is rather sensitive to the choice of baryon radius $r_B$.
For example, for $r_B = 0.35$~fm there is a narrower minimum around $T \simeq 173$~MeV in the temperature dependence of the $\chi^2$ while for $r_B = 0.25$~fm there are two local minima, at $T \simeq 159$~MeV and at $T \simeq 234$~MeV.

Inclusion of a bag-like eigenvolume with $r_p = 0.5$~fm (second EV case) leads to even more dramatic changes in the $\chi^2$ profile.
It is seen from Fig.~\ref{fig:chi2-vs-T} that the temperature dependence of the $\chi^2$ has a rather complicated two-minimum structure, and the data is described better than at the best fit in the point-particle ($r_p = 0$~fm) case in a very wide $170-320$ MeV temperature range. The global minimum is located at $T = 274$~MeV with $\chi^2/N_{\rm dof} \simeq 15.1/10$. 
It is interesting that this temperature is very close to the $T_c \simeq 270$~MeV temperature of the first-order phase transition in the pure SU(3) Yang-Mills theory~\cite{lQCDsu3}.
It has been argued that this theory is relevant for the description of the evolution of the initially pure glue matter created in the heavy-ion collisions~\cite{PureGlue}.
We also note that, at the global minimum, the total system volume (radius) 
per unit slice of rapidity,
decreases somewhat, from $R = 10.8$~fm at $r_p=0$~fm to $R = 9.8$~fm at $r_p = 0.5$~fm.
In general, the inclusion of eigenvolume leads to the increase of volume at the same temperature compared to point-particle case. In our case, however, the temperature is essentially larger for $r_p = 0.5$~fm case and for this reason the volume is decreased.
The same picture is obtained if the finite width of the resonances is neglected. The bag model inspired parametrization yields by far the strongest effect on thermal fits.

Finally, the third case, where only baryon-baryon and antibaryon-antibaryon eigenvolume interactions are considered, yields a rather wide minimum around $T = 172$~MeV with minimum $\chi^2 /N_{\rm dof} \simeq 24.4/10$, also yielding an improvement versus the point-particle HRG. The result is rather similar to the first EV case, and, likewise, this model is fully consistent with lattice QCD pressure up to at least $T=200$~MeV.

\begin{figure}[!t]
\centering
\includegraphics[width=0.49\textwidth]{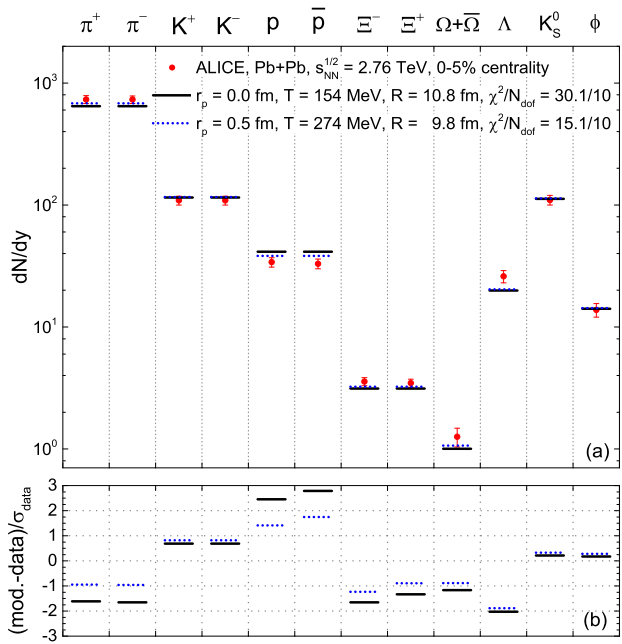}
\caption[]{(Color online)
(a) Hadron yields in the 0-5\% most central Pb+Pb collisions at 2.76~TeV measured by the ALICE collaboration, and calculated within the point-particle HRG model (solid black curve) and the bag-like eigenvolume HRG model (dotted blue line). In the latter case the bag-like constant is fixed in order to reproduce the hard-core proton radius of 0.5 fm.
(b) The normalized deviation of the hadron yields calculated within the HRG from the ALICE data.
}\label{fig:yields}
\end{figure}

Let us now focus on the bag-like eigenvolume where the effect on thermal fits was found to be by far the strongest.
Figure~\ref{fig:yields} shows the comparison between the ALICE data and the HRG model fit to the hadron yields for the point-particle and the bag-like cases. It is seen that the eigenvolume model describes considerably better the yields of pions and protons than the point-like HRG, while the quality of description of other yields remains very similar.
In general, the deviations of all considered yields from the data does not exceed 2$\sigma$ for $r_p = 0.5$~fm.
The improvement of the description of the low $p/\pi$ ratio here is due to the eigenvolume effects only, and is not related to other proposed explanations, such as the incomplete hadron spectrum~\cite{SABR2014,Nor10}, the chemical non-equilibrium at freeze-out~\cite{SHARE,Begun,VovchenkoEntropy}, or the post-freeze-out hadronic reactions~\cite{Becattini2014,Becattini2013}.
It will be interesting to see whether similarly good description at high temperatures within the bag-like EV model will be found for the hadron yield data from the ongoing ALICE heavy-ion run at $\sqrt{s_{\rm NN}} = 5.02$~TeV.

\begin{figure}[!t]
\centering
\includegraphics[width=0.49\textwidth]{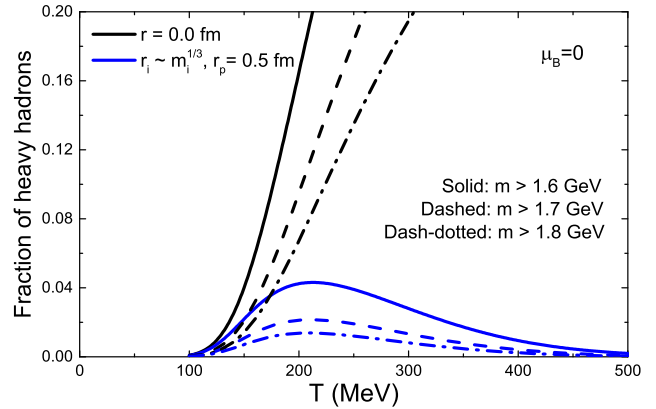}
\caption[]{(Color online)
The temperature dependence of the relative contribution of the heavy hadrons with different lower mass cutoffs to the total density of all hadrons for the point-particle HRG (black lines) and for the bag-like HRG with $r_p = 0.5$~fm (blue lines).
}\label{fig:heavy}
\end{figure}

In our calculations we have included hadrons with masses only up to $N(2600)$.
It would seem that at temperatures beyond 200~MeV the more heavier resonances will play a major role and that they cannot be neglected. Note, however, that in the bag-like EV model the eigenvolume of the hadron is proportional to its mass, thus, the heavier hadrons are more strongly suppressed than the lighter ones. In Figure~\ref{fig:heavy} the temperature dependence of the relative contribution of the heavy hadrons with different lower mass cutoffs to the total density of all hadrons is shown for the point-particle HRG and for the bag-like HRG with $r_p = 0.5$~fm.
For the point-particle case a rapid increase of the heavy hadrons fraction is seen above $T=160$~MeV, showing signs of Hagedorn divergence. A similar picture would emerge for HRG with hadrons of finite but identical eigenvolume.
For the bag-like case the role of heavy hadrons is clearly suppressed at high temperatures due to their large eigenvolumes, and their fraction never exceeds the one at $T=155$~MeV for the point-particle HRG. From this we conclude that the uncertainties introduced by using the incomplete spectrum of heavy resonances in bag-like HRG at high temperatures does not exceed ones for the point-particle HRG at $T=155$~MeV.

The non-monotonic behavior of the heavy hadrons fraction seen in Fig.~\ref{fig:heavy} also hints at the origin of the double-minimum structure in the $\chi^2$ temperature dependence depicted in Fig.~\ref{fig:chi2-vs-T}. 
Indeed, different magnitudes of eigenvolume suppression for different hadrons leads to non-monotonic temperature dependence of many hadron yield ratios.
This concerns particularly the baryon-to-meson ratios: larger baryons are suppressed more strongly at high temperatures/densities than the smaller-sized mesons.
This implies that a measurement of a particular hadron yield ratio may correspond to two different temperatures in the eigenvolume models~(see also Refs.~\cite{Vovchenko:2016eby} and \cite{Satarov:2016peb} for a systematic analysis of this phenomenon).

In order to additionally check the sensitivity of the results to the choice of the hadron spectrum, 
we have performed the same fit procedure but including only the hadrons with masses up to the mass of the $\Omega$ baryon. 
We have also performed the analysis using the particle table from THERMUS-3.0,
which also includes charm and the light nuclei.
In particular we have checked the influence of including the deuteron yield into the thermal fit.
Additionally, we studied influence of the inclusion of 
the $\sigma$ ($f_0(500)$) and
$\kappa$ ($K_0^*(800)$) mesons.  
It has been pointed out that their inclusion into HRG should be treated with care~\cite{Broniowski:2015oha,Pelaez:2015qba}.
In all cases we have obtained similar $\chi^2$ profile to the one shown in the Fig.~\ref{fig:chi2-vs-T}: for the point-particle HRG the $\chi^2$ is minimized at $T\simeq155$~MeV,
while the finite eigenvolume with $r_p=0.5$~fm gives a better fit quality in a wide $170-320$ MeV temperature range.

The bag-like ($r_p = 0.5$~fm) fit is characterized by 
rather high values of the
packing fraction $\eta$ (fraction of total volume occupied by hadrons of finite size),
which quantifies the role of the eigenvolume effects.
At $T=270$~MeV the packing fraction is equal to $\eta \simeq 0.15$,
a value where van der Waals equation already deviates noticeably from the equation of state of hard spheres. While we do not require hadrons to behave exactly like non-deformable hard spheres this does mean that at such high packing fractions the behavior of thermodynamic quantities may depend strongly on the way eigenvolume interactions are modeled (see, e.g., Fig.~1 in Ref.~\cite{mf-2014}),
and that the high-temperature part may be outside the range where
van der Waals excluded volume model can be applied safely.
In particular, the high-temperature part of the fit is plagued by the superluminal behavior of the speed of sound, namely $c_s^2 \sim 1$ for $T>250$~MeV.
The superluminal speed of sound is a known problem of the EV model,
and avoiding it completely would require a modification of the model.
For these reasons it can also be useful to consider more complicated formulations of the EV model in order to check the robustness of the results at high ($T>250$~MeV) temperatures.
These, for instance, can either treat 
differently the correlations between hadrons of different size~\cite{GKK,BugaevEV}, or go beyond the van der Waals low-density extrapolation, thus, avoiding the acausality problem until much higher temperatures~\cite{mf-2014,mf-1992,SBM}. While we expect the formulation given by Eqs.~\eqref{eq:Pex} and \eqref{eq:nex} to capture all the essential features of the multi-component eigenvolume gas, a more precise study on this subject could require a careful examination of the different eigenvolume models. We note that the problems mentioned above do not appear in the other two scenarios for eigenvolume interactions which has been considered in the present paper.

In all our fits we did not include yields of the light nuclei. 
There are reasons why inclusion of light nuclei into thermal fits is questionable, in particular related to their small binding energies. 
Nevertheless, they are sometimes included into thermal fits~\cite{SABR2014}, and may help to stabilize them in some cases. 
The inclusion of light nuclei was checked not to influence the fits considerably for the mass-proportional EV parameterization.
For another two EV scenarios it was found that the fit is extremely sensitive to the assumptions regarding the eigenvolumes of different nuclei. 
For instance, the fit changes drastically when eigenvolume of a deuteron is changed from the size of a single proton ($v_d = v_p$) to the size of two protons~($v_d = 2 v_p$, see~\cite{Vovchenko:2016mwg}). Due to this large sensitivity, and to the generally questionable nature of inclusion of light nuclei into thermal fits, the addition of light nuclei into fits is not considered further.

In the present work we do not attempt to pinpoint the accurate value of the chemical freeze-out temperature, or to determine the hadron eigenvolumes. 
To the contrary, the results presented here indicate that one cannot extract the ``hadronization temperature'' (or the chemical freeze-out temperature) with very high reliability from the LHC heavy-ion data, at least not until the role of the eigenvolume interactions is properly clarified. In particular,
the high temperature minimum given by the bag-like model
should not be interpreted as evidence
for chemical freeze-out at 300~MeV.
It can be interesting to perform a similar analysis on the data at lower energies, such as RHIC and SPS. There the finite (baryo)chemical potential will play an additional role. According to our preliminary findings the thermal fits at lower energies are similarly very sensitive to the modeling of the eigenvolume interactions~\cite{VS-SPS-fits}. It is evident that proper constraints on eigenvolume interactions for all the different hadron pairs are needed.

\section{Summary}

In summary, the data of the ALICE collaboration on the mid-rapidity yields of hadrons in central Pb+Pb collisions at $\sqrt{s_{\rm NN}} = 2.76$~TeV is analyzed within the HRG model 
with excluded volume effects using three drastically different scenarios. 
It is found that a two-component eigenvolume HRG model with point-like mesons and baryons with hard-core radius of $r_B = 0.3$~fm describes the data in a rather wide temperature range of 155-210~MeV, with fit quality comparable to the point-particle case.
The eigenvolume model with the mass-proportional eigenvolumes 
($r_i \sim m_i^{1/3}$), fixed to a proton hard-core radius of 0.5~fm, describes the data better than the conventional point-particle HRG model in a very wide temperature range of 170-320 MeV. 
Finally, the model where only baryon-baryon and antibaryon-antibaryon eigenvolume interactions are included, with $r_B$ fixed by previous fit to nuclear ground state, yields a wide minimum around $T = 172$~MeV.

The obtained results show that the thermal fits are a very delicate procedure, which is surprisingly sensitive
to the modeling of the eigenvolume interactions. It appears that, at the moment, the chemical freeze-out temperature
can be determined from the LHC heavy-ion data on hadron yields only with a sizable uncertainty,
at least in the case when full chemical equilibrium is assumed.
Evidently, the modeling of eigenvolume interactions in HRG needs to be properly clarified in order to address this problem.
It is predicted that a similar picture emerges at the higher collision energies, and, thus, needs to be taken into account in the future analysis
of the data which will be obtained during the ongoing ALICE heavy-ion run at 5.02 TeV.

\vspace{0.5cm}
\noindent
{\it Acknowledgements.}~~ We are thankful
to P.~Alba, A.~Andronic, P.~Braun-Munzinger, J.~Cleymans, M.I.~Gorenstein, J.~Rafelski, Nan Su, and A.~Tawfik for fruitful comments and discussions.
This work was supported by HIC for FAIR within the LOEWE program of the State of Hesse.

\end{document}